\documentclass[%
 reprint,
superscriptaddress,
 amsmath,amssymb,
 aps,
]{revtex4-1}

\usepackage{graphicx}% Include figure files
\usepackage{dcolumn}% Align table columns on decimal point
\usepackage{bm}% bold math
\usepackage{amsmath}
\begin{document}

%\preprint{APS/123-QED}

\title{Suppression of electron thermal conduction by whistler turbulence in a sustained thermal gradient}

\author{G. T. Roberg-Clark}
\email{grc@umd.edu}
\affiliation{Department of Physics, University of Maryland College Park, College Park, MD 20740, USA}
\author{J. F. Drake}%
\email{drake@umd.edu}
\affiliation{Department of Physics, University of Maryland College Park, College Park, MD 20740, USA}
\affiliation{Institute for Physical Science and Technology, University of Maryland, College Park, MD 20742, USA}
\affiliation{Institute for Research in Electronics and Applied Physics, University of Maryland, College Park, MD 20742, USA}
\affiliation{Joint Space-Science Institute (JSI), College Park, MD 20742, USA}
\author{C. S. Reynolds}%
\email{chris@astro.umd.edu}
\affiliation{Department of Astronomy, University of Maryland College Park, College Park, MD 20740, USA}
\affiliation{Joint Space-Science Institute (JSI), College Park, MD 20742, USA}
\author{M. Swisdak}%
\email{swisdak@umd.edu}
\affiliation{Department of Physics, University of Maryland College Park, College Park, MD 20740, USA}
\affiliation{Institute for Research in Electronics and Applied Physics, University of Maryland, College Park, MD 20742, USA}
\affiliation{Joint Space-Science Institute (JSI), College Park, MD 20742, USA}

% \altaffilmark{1}

%\altaffiltext{1}{Department of Physics, University of Maryland, College Park, College Park, MD 20740}%

%\altaffiltext{2}{Department of Astronomy, University of Maryland, College Park, College Park, MD 20740}

\date{\today}


\begin{abstract}
The dynamics of weakly magnetized collisionless plasmas in the presence of an imposed temperature gradient along an ambient magnetic field is explored with particle-in-cell simulations and modeling. Two thermal reservoirs at different temperatures drive an electron heat flux that destabilizes off-angle whistler-type modes. The whistlers grow to large amplitude, $\delta B / B_{0} \simeq 1$, and resonantly scatter the electrons, significantly reducing the heat flux. A surprise is that the resulting steady state heat flux is largely independent of the thermal gradient. The rate of thermal conduction is instead controlled by the finite propagation speed of the whistlers, which act as mobile scattering centers that convect the thermal energy of the hot reservoir. The results are relevant to thermal transport in high $\beta$ astrophysical plasmas such as hot accretion flows and the intracluster medium of galaxy clusters.

%\begin{description}
%\end{description}
\end{abstract}

\maketitle

%\tableofcontents

\textit{Introduction.} Thermal conduction is integral
to a wide variety of phenomena occurring in space and astrophysical
plasmas. Ascertaining the rate of thermal conduction in such systems
is therefore of fundamental importance. In particular, the
microphysics of weakly collisional, weakly magnetized plasmas, which
is not fully understood, may play a pivotal role in determining the
transport properties of the global system \cite{Rincon2014}. A
magnetic field makes thermal conduction anisotropic and when the
plasma $\beta$ is greater than order unity, the system is susceptible
to a host of microscale kinetic instabilities which tend to suppress
thermal fluxes via particle scattering
\cite{Ramani1978,Levinson1992,Gary2000,Kunz2014,Komarov2016,Riquelme2016}. Such
instabilities are expected to operate in rarefied plasma environments
such as the Intracluster Medium (ICM) of galaxy clusters
\cite{Schekochihin2010,Kunz2011,Roberg-Clark2016} as well as hot
accretion flows \cite{Quataert1997,Sharma2007} and the solar wind
\cite{Hellinger2013,Hollweg1974,Gary1994}.

The impact of instabilities on transport is tied to their nonlinear
evolution, which in turn is influenced by the input of free energy
from the surrounding astrophysical environment that can be included
through appropriate boundary conditions in numerical models. Examples
include those used in shearing-box \cite{Kunz2016,Riquelme2016} and
compressing-box \cite{Sironi2015a,Sironi2015b} simulations of
instabilities driven by pressure anisotropies and their impact on the
transport of heat and momentum in accretion flows. Here we focus
on the dynamics of a system in contact with two thermal
reservoirs at different temperatures, driving a heat flux parallel to
an ambient magnetic field. We use particle-in-cell (PIC) simulations
to model the resulting heat flux instability and steady-state
suppressed thermal conduction with a sustained thermal gradient. This
model is in contrast to previous work in which the heat flux
instability was studied as an initial value problem
\cite{Roberg-Clark2016}.

\begin{figure*}
    \centering
    \includegraphics[scale=.715]{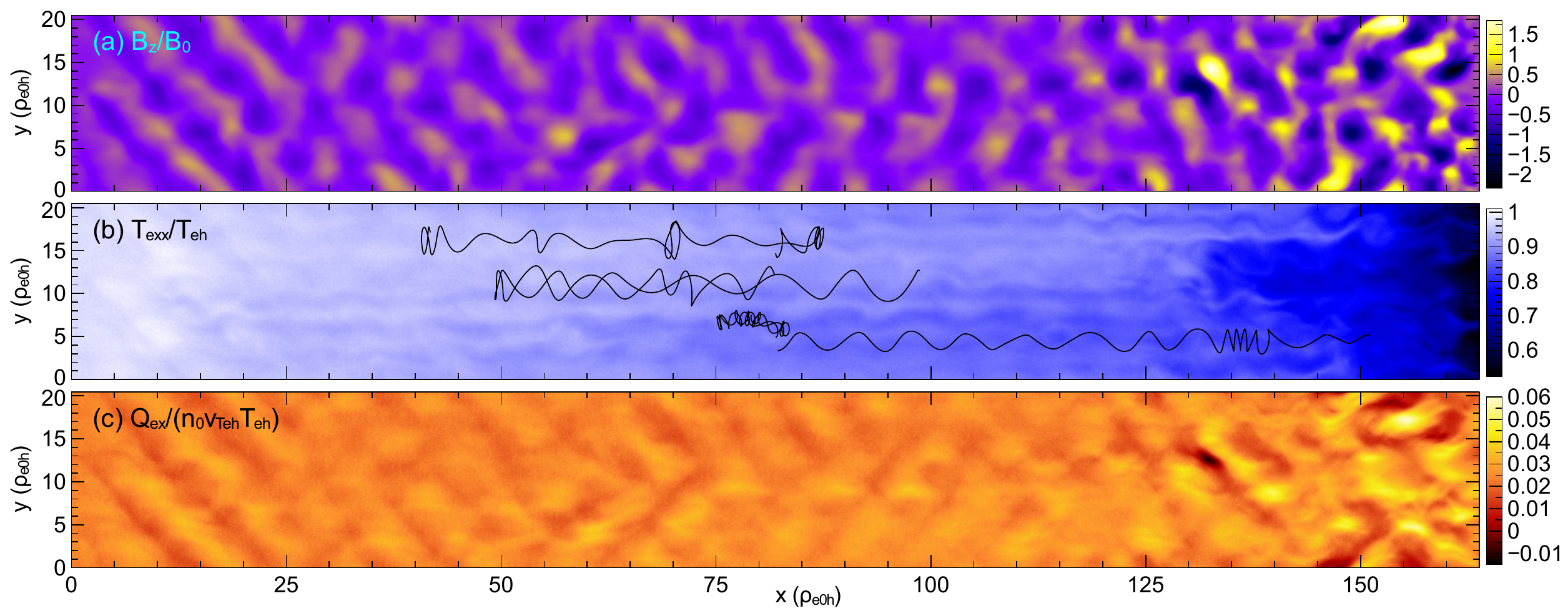}
    \caption{Two dimensional plots from the largest simulation ($L_{x}=2L_{0}$) in a saturated state at $t=800 \: \Omega_{e0}^{-1}$. (a) Fluctuations in out-of-plane $B_{z}$ (b) Temperature $T_{exx}$ with four self-consistent particle trajectories overlaid. (c) Heat flux $q_{ex}$. }
    \label{fig:1}
\end{figure*}

\textit{Numerical Scheme.} We carry out two-dimensional (2D)
simulations using the PIC code $\tt{p3d}$ \cite{Zeiler2002} to model
thermal conduction along an imposed temperature gradient in a
magnetized, collisionless plasma with open boundaries. $\tt{p3d}$
calculates particle trajectories using the relativistic Newton-Lorentz
equations and the electromagnetic fields are advanced using Maxwell's
equations. The ends of the simulation domain act as thermal reservoirs
at two different temperatures $T_{h} > T_{c}$ separated by a distance
$L_{x}$, forming a temperature gradient $T' \equiv
(T_{h}-T_{c})/L_{x}$ and driving a heat flux. An initially uniform
magnetic field $\mathbf{B_{0}}=B_{0} \mathbf{\hat{x}}$ threads the
plasma along the gradient and is free to evolve in time. The initial
particle distribution function is chosen to model the free-streaming
of particles from each thermal reservoir and has the form \small
\begin{multline}\label{eqn:1}
f(\mathbf{v},t=0) = f_{h} + f_{c} \\ = \frac{n_{0}}{\pi^{3/2}} \left( \frac{ e^{-v^2/v_{Th}^2}}{v_{Th}^3}\theta(v_{\parallel}) + \frac{e^{-[(v_{\parallel} + v_{d})^2 + v_{\perp}^2]/v_{Tc}^2}}{v_{Tc}^3(1+\text{erf}(v_{d}/v_{Tc}))}\theta(-v_{\parallel}) \right)
\end{multline}
\normalsize where $n_0$ is the initial density, $\theta$ is the
Heaviside step function, $v_{T}=\sqrt{2T/m}$ is the thermal speed, and
the parallel and perpendicular directions are with respect to
$\mathbf{B_0}$. The cold particles are given a parallel drift speed
$v_{d}$ to ensure zero net current ($\langle v_{\parallel} \rangle =
0$) in the initial state while the error function
$\text{erf}(v_{d}/v_{Tc})$ makes the density of hot and cold particles
equal. $f_0$ also has nonzero pressure anisotropy ($\langle
v_{\parallel}^2 \rangle \ne \langle v_{\perp}^{2}/2 \rangle$) and a
heat flux $q_{\parallel} =\langle v_{\parallel} v^2 \rangle =
q_{0}$. $f_0$ is not unstable in a 1D system since only off-angle
modes resonate with particles near the large phase space
discontinuity in $f_0$ at $v_{\parallel}=0$ (Fig. S1a).

When particles exit the open boundaries they are re-injected with
velocities pulled from $f_h$ (at $x=0$) or $f_c$ ($x=L_{x}$). The
drift velocity $v_{d}$ is then recalculated at each time step to
ensure that the current of re-injected particles cancels the current
of outgoing particles at the cold reservoir. The electromagnetic field
components at the thermal reservoir boundaries are $F_{y}=0,\:
\partial{F_{x}}/\partial{x} = \partial{F_{z}}/\partial{x}=0$ where
$F=(\mathbf{E},\mathbf{B})$. Periodic boundary conditions are used for
both particles and fields in the $y$ direction. Ions in the simulation are not evolved in time and act
as a charge-neutralizing background. The
subscript $e$ denotes an electron quantity.
\begin{center}
\begin{tabular}{ |p{2cm}|p{1cm}|p{1.25cm}|p{2.75cm}| }
 \hline
 \multicolumn{4}{|c|}{Table 1} \\
 \hline
 $L_x$ & $\beta_{e0h}$& $T_{ec}/T_{eh}$ & $q_{ex,f} / (n_{0}v_{p}T_{eh})$ \\
 \hline
 $L_{0}= 82 \: \rho_{e0h}$ & $64$ & $1/2$ & $3.44$\\
 $L_{0}/2$ & $64$ & $1/2$ & $3.30$\\
 $2L_{0}$ & $64$ & $1/2$ & $3.26$\\
 $L_{0}$ & $32$ & $1/2$ & $3.46$\\
 $L_{0}$ & $128$ & $1/2$ & $3.19$ \\
 $L_{0}$ & $64$  & $1/4$ & $2.56$\\
 \hline

\end{tabular}
\end{center}

\textit{Simulation Parameters.} We have performed six simulations in which $L_{x}$, $B_0$ and $T_{ec}/T_{eh}$ are varied independently so as to change $T'$ and $\beta_{e0h}=4\pi n_{0}T_{eh}/(B_{0}^{2}/2)$. The baseline simulation has $L_x=L_0=82 \: \rho_{e0h}$, $\beta_{e0h}=64$, $T_{ec}=T_{eh}/2$, $\omega_{pe}/\Omega_{e}=40$, and $T_{eh}/(m_{e}c^2)=.02$, where $\rho_{e0h}=v_{Teh}/\Omega_{e0}$ is the gyroradius, $\Omega_{e0} = eB_{0}/(m_{e}c)$ is the cyclotron frequency, and $\omega_{pe}=(4\pi n_{0}e^{2}/m_{e})^{1/2}$ is the plasma frequency. The parameters for each simulation are listed in Table 1. Each simulation uses 560 particles per cell, has a transverse length $L_{y}$ of $20 \: \rho_{e0h}$, and is run to $t=800 \: \Omega_{e0}^{-1}$. The largest simulation ($L_x= 2L_0$) has a spatial domain of $32768$ by $4096$ grid cells.

\textit{Whistler Turbulence.}  Initializing the
simulations with $f_0$ leads to an impulse of transient fluctuations
in the out-of-plane magnetic field $B_{z}$ that propagate towards the
hot thermal reservoir (evidence for this is shown later). These
fluctuations are driven by the initial pressure anisotropy and quickly
lead to a sharp drop in the anisotropy to the marginally stable level
for firehose-type modes (not shown). The fluctuations rapidly damp and
become dynamically unimportant in the simulations and are not
discussed further.

The reinjection and mixing of hot and cold particles results in a
continuous source of heat flux in the simulation domain. The heat flux drives
off-angle ($k_{y} \simeq k_{x}$), slowly propagating ($\omega/k \ll
v_{Teh})$, elliptically polarized whistler modes that reach large
amplitude, $\delta B/B_{0} \simeq 1$ (fig. \ref{fig:1}a), and strongly
scatter electrons, isotropizing the electron distribution function
(see the supplementary material). The heat flux $q_{ex}$ drops well
below its initial value $q_{ex0}$. Some reflection of waves occurs at
the cold plate boundary but the heat flux is insensitive to the length
of the simulation domain, confirming that such reflection does not
impact the integrated results.

Strong scattering by the whistlers causes inherently 2D
structures to develop in quantities such as the temperature $T_{exx} =
m_e\langle v_x^{2} \rangle$ (figure \ref{fig:1}b) and heat flux
$q_{ex}$ (figure \ref{fig:1}c). In figure \ref{fig:1}b the
trajectories of four electron macro-particles from the simulation,
tracked starting from an initial position $x=L_{x}/2$ for a period of
$87.5 \: \Omega_{e0}^{-1}$ in steady state, are overlaid over
$T_{exx}$, which does not vary appreciably during the time of the
orbits. Some particles reverse their parallel velocity several times
as a result of scattering in the strong magnetic fluctuations. Because
the system is 2D the particle out-of-plane
canonical momentum, $p_{ez}=m_{e}v_{z} -eA_{z}$, is a conserved
quantity. Since $A_{z} \sim yB_{x}$ and kinetic energy is mostly
conserved in the magnetic fluctuations, the electrons are confined to
relatively narrow channels in $y$.

\begin{figure}
    \centering
    \includegraphics[scale=.43]{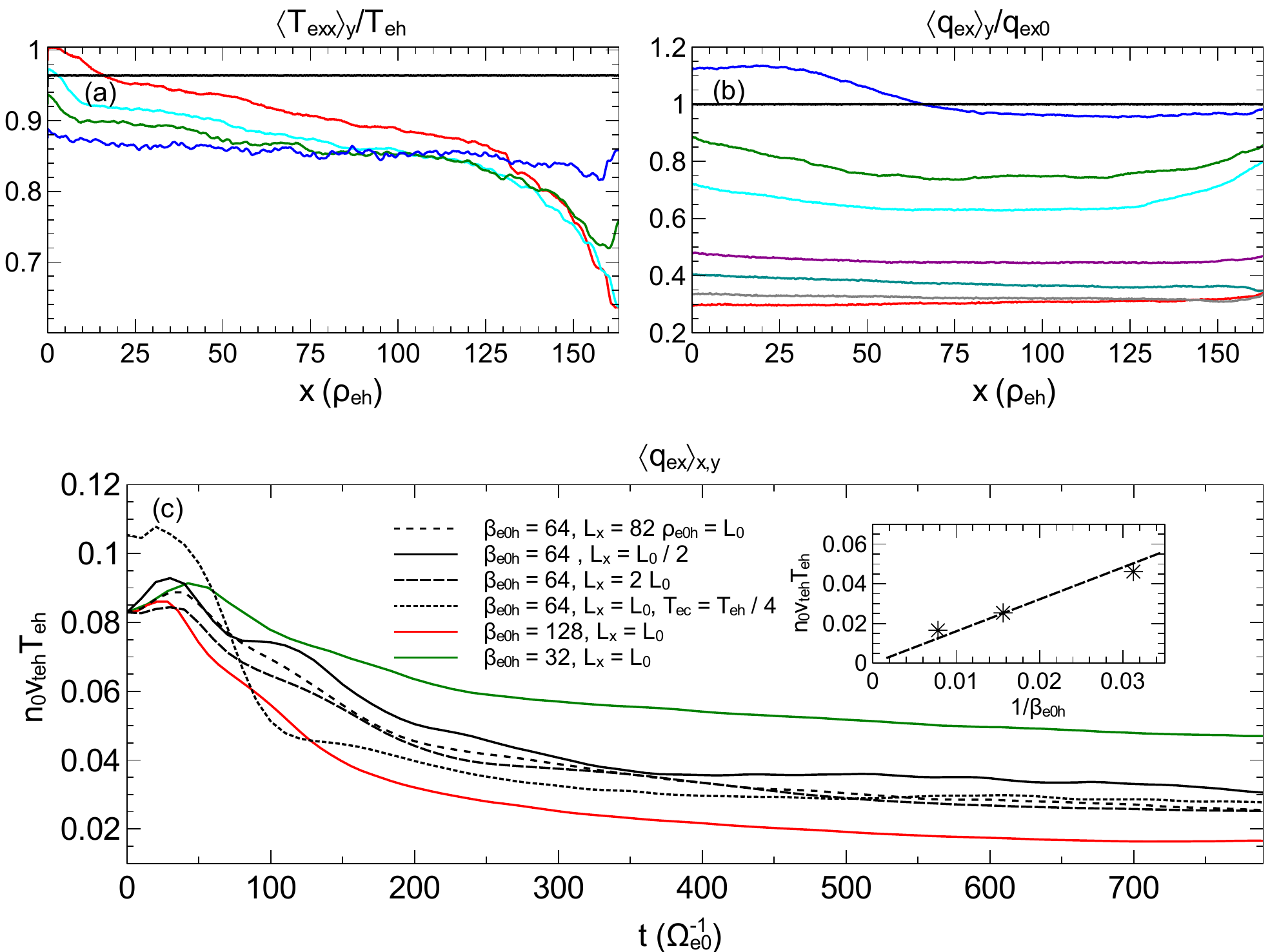}
    \caption{Temperature and heat flux profiles. (a) Line plots of $y$-averaged $T_{exx}$ for the run with $L_{x}=2L_{0}$ at times $t\Omega_{e0}=0$ (black), $80$ (blue), $200$ (green), $400$ (cyan), and $800$ (red). (b) Line plots of $y$-averaged heat flux at times $t\Omega_{e0}^{-1}=0$ (black), $30$ (blue), $100$ (green), $150$ (cyan), $300$ (purple), $450$ (grey-blue), $600$ (grey), and $800$ (red). (c) Box-averaged $\langle q_{ex}(t) \rangle_{x,y}$ for the six simulations in Table 1. \textit{Inset}: Linear fit to the steady state heat flux $q_{ex,f}$ as a function of $1/\beta_{e0h}$. }
    \label{fig:2}
\end{figure}

\textit{Suppression of Thermal Conduction.}  Suppression of the heat flux
develops over a time of hundreds of $\Omega_{e0}^{-1}$ resulting in a
steady state in which a continuous temperature profile has formed
between the hot and cold reservoirs (fig. \ref{fig:2}a) and the heat
flux has leveled off to a nearly constant value
(fig. \ref{fig:2}b). Fig. \ref{fig:2}c shows the time profiles of
average heat flux $\langle q_{ex} \rangle_{x,y}$ for six simulations.
The expectation for a system subject to Coulomb scattering
(or another scattering process) is that the heat flux is diffusive,
$\mathbf{q}_{e} \propto - \nabla{T_{e}}$. We find instead that the final heat
flux is insensitive to the ambient gradient. The black lines in fig. \ref{fig:2}c correspond to simulations
with a fixed $\beta_{e0h}=64$ but differing box lengths or hot to cold
temperature jumps. For all of these runs the heat flux settles at
around $0.03 \: n_{0}T_{eh}v_{Teh}$ . Thus, the heat flux rather than the gradient controls the dynamics. As long as $T_{eh}$ is significantly greater than $T_{ec}$, the
hot plate controls the final heat flux. However, the two simulations with differing $\beta_{e0h}$ have noticeably different asymptotic heat fluxes that follow
the scaling $\langle q_{ex} \rangle_{x,y} \propto 1/\beta_{e0h}$ (fig \ref{fig:2}c inset).  To explain this result we turn to the physics of scattering by
large-amplitude whistler waves.

\textit{Scattering by Whistlers.} The physics of resonant interactions
of particles with elliptically polarized whistlers is well-documented
in the literature (see e.g. \cite{Roberg-Clark2016} and references
therein). In the frame of a single off-angle whistler wave, total
kinetic energy is conserved and particles which satisfy the various
resonance criteria $v_{\parallel} = n\Omega_{0}/k, n=0,\pm 1, \pm2,
...$, are trapped \cite{Karimabadi1992a}. For $\delta B/B_{0} \ll 1$,
resonant particles experience small oscillations in the
$v_{\parallel}/v_\perp$ plane. For large-amplitude whistlers ($\delta
B/B_{0} \gtrsim 0.3$) resonances can overlap, leading to irreversible
diffusive behavior along circular, constant energy curves in the
whistler wave frame \cite{Karimabadi1992a}. In the presence of
multiple whistlers with differing parallel phase speeds some diffusion
may also occur perpendicular to circles of constant energy
\cite{Karimabadi1992a}. Resonance overlap is an effective mechanism
for heat flux suppression since it causes large deflections in the particle
pitch angle $\phi = \tan^{-1}(v_{\perp}/v_{\parallel})$, quenching the
parallel heat flux \cite{Roberg-Clark2016}.

\begin{figure}
    \centering
    \includegraphics[scale=.57]{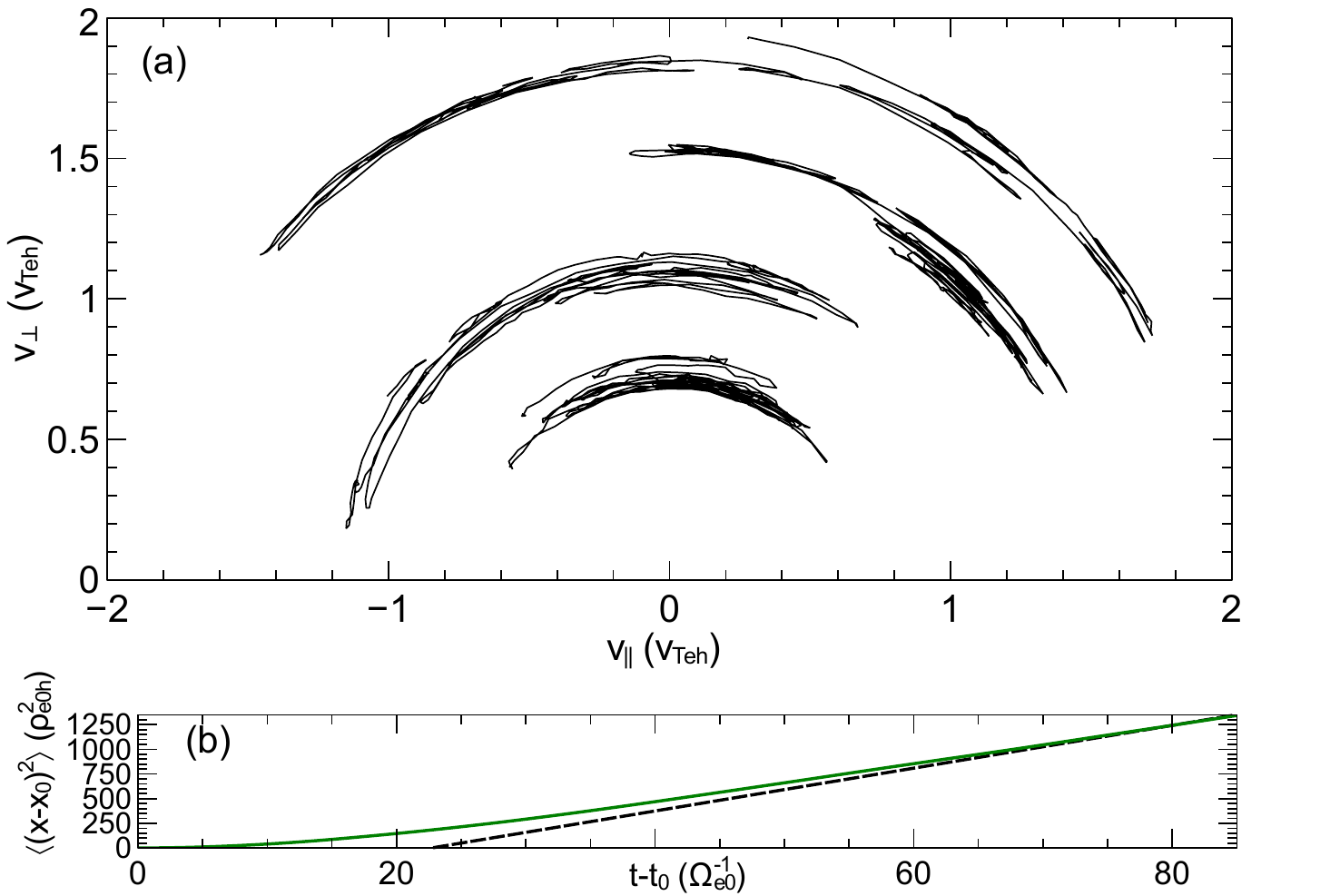}
    \caption{Evidence for scattering of electrons by whistlers. (a) Trajectories of particles in $v_{\parallel}$,$v_{\perp}$ space showing significant deflection of pitch angle $\phi = \tan^{-1}(v_{\perp}/v_{\parallel})$. (b) Plot of $\langle (x(t)-x(t_{0}))^{2} \rangle$ with linear fit to the slope representing diffusion coefficient $D$. }
    \label{fig:3}
\end{figure}

To demonstrate that this is the physics at play in our simulations, in
fig. \ref{fig:3}a we show a resonance diagram in
$v_{\parallel}-v_{\perp}$ for four trapped particles with differing
energy in the simulation with $L=2L_{0}$ at steady state. Particle
energy is mostly conserved and the primary diffusion is in pitch angle
\cite{Karimabadi1992a}. All the particles display significant
deflection so the bulk of particles undergo trapping by the
whistlers. Also of note is that the nearly-circular contours in
velocity space are effectively centered about $v_{\parallel}=0$,
indicating that the whistler phase speed is small compared to the
thermal speed $v_{Teh}$.
\begin{figure}
    \centering
    \includegraphics[scale=.35]{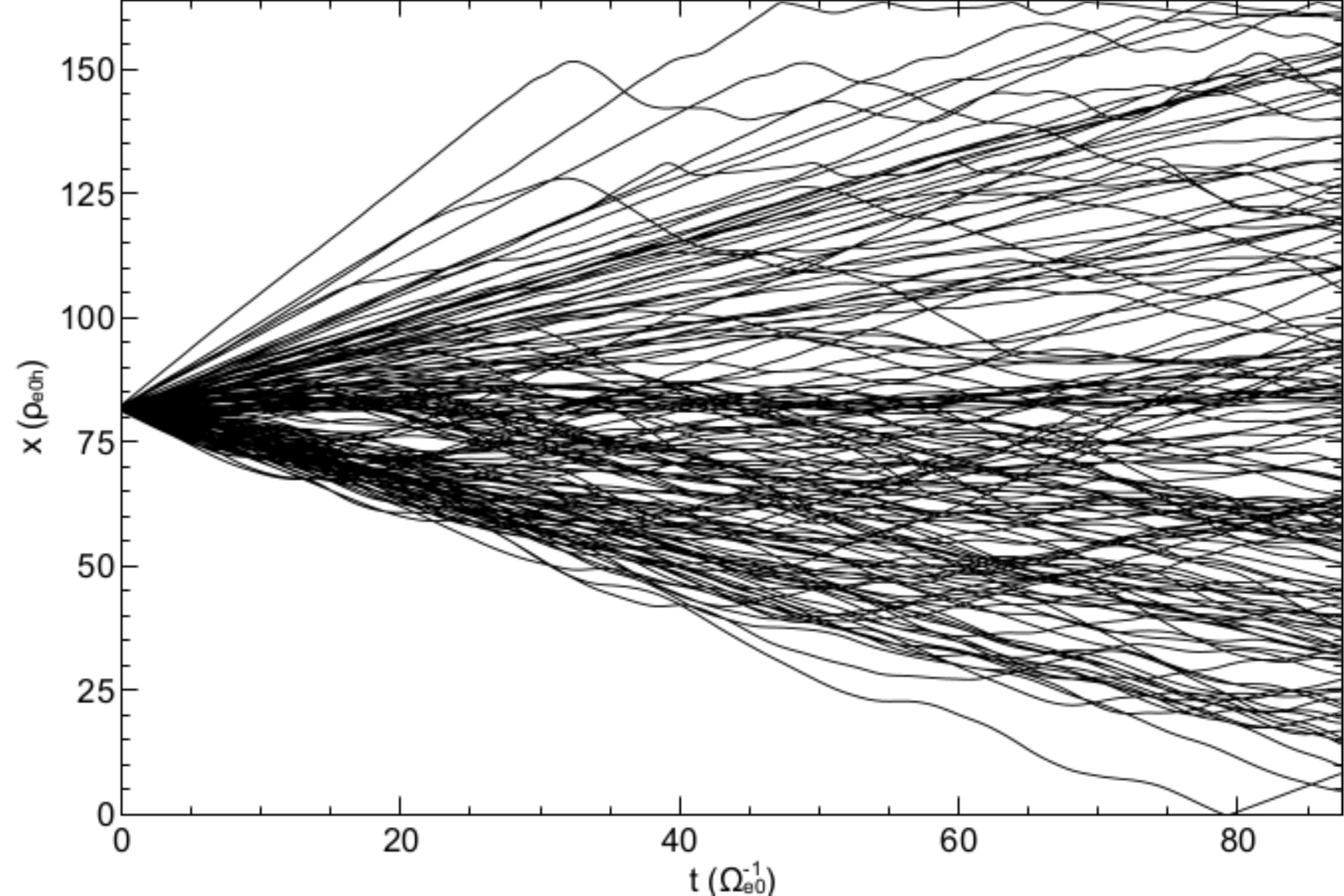}
    \caption{Line plots of $x(t)$ for $150$ particles in the $L_{x}=2L_{0}$ simulation indicating diffusive behavior.}
    \label{fig:4}
\end{figure}

To quantify the rate of scattering by the whistlers we calculate the
quantity $\langle (x(t)-x(t_{0}))^{2}\rangle$ by averaging over
individual trajectories of roughly $8000$ particles for the
$L_{x}=2L_{0}$ simulation (fig. \ref{fig:3}b). The diffusion rate
$D=\langle v^2 \rangle \tau$ is half the linear slope of
$(x-x_{0})^{2}$ at late time, where $\tau$ is the scattering time. We
find $\tau \simeq 6.80 \: \Omega_{e0}^{-1}$. We plot $x$ versus time
for $150$ particles in fig. \ref{fig:4} to illustrate the particle
motion. Some particles are diverted back towards the initial particle
location at $y=L_{0}$ once scattering becomes significant while
others maintain their initial direction of propagation. The linear
trend of mean-squared displacement in \ref{fig:3}b is evidence for
diffusive behavior. Pitch angle scattering in a spectrum of whistler
turbulence was also reported by \cite{Keenan2016}.

\textit{Steady State Heat Flux.} The results of fig. \ref{fig:2}a have
demonstrated that the asymptotic rate of thermal conduction in the
presence of large-amplitude whistler waves is largely independent of
the temperature gradient and instead follows a scaling $1/\beta_{e0h}
$. A simple explanation for this result, consistent with a comment in \cite{Levinson1992}, is that whistlers act as particle
scattering centers that propagate at their phase speed
$v_{p}=\omega/k$ and control the net flow of high-energy particles
carrying the bulk of the heat flux. The resulting heat flux is simply
the product of the phase speed and the thermal energy of the hot
plasma, $q_{ex}\sim n_0v_{p}T_{eh}$.

The whistler wave phase speed is determined via the cold plasma dispersion relation, $\omega = k^{2}\rho_{e}^2 \Omega_{e}/\beta_{e}$. Taking $k \rho_{e} \sim 1$ (as in \cite{Roberg-Clark2016}) for whistlers at high $\beta_{e}$, we find
\begin{equation} \label{eqn:2}
\frac{\omega}{k} \sim \frac{v_{Te}}{\beta_{e}}.
\end{equation}
\begin{figure}
    \centering
    \includegraphics[scale=.54]{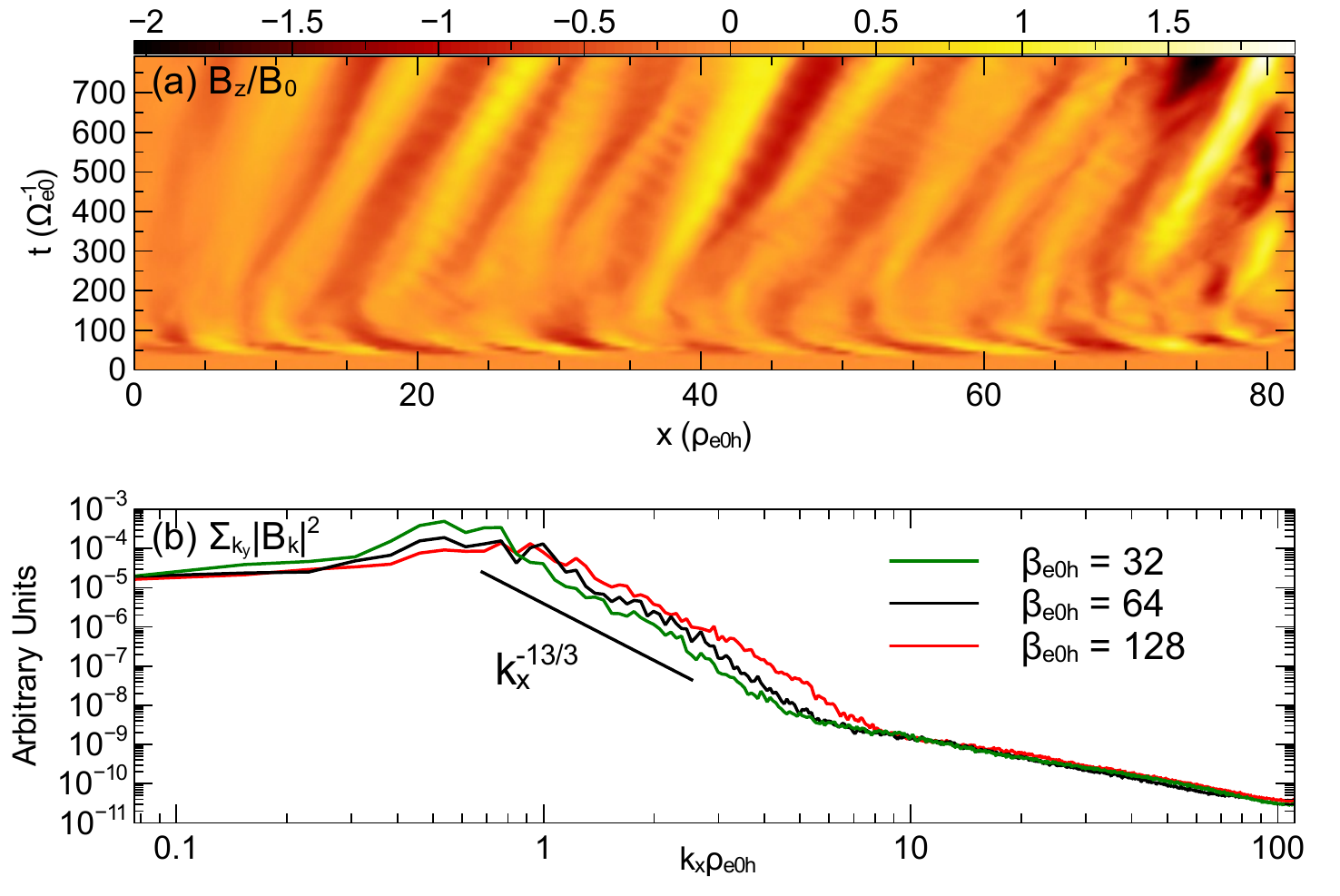}
    \caption{(a) Spacetime plot ($t$ versus $x$) of whistler fluctuations propagating through the simulation with $L_{x}=L_{0},\beta_{e0h}=64$ at $L_{y}/2$. At early times the initial condition $f_{0}$ produces fluctuations that reverse direction and are overtaken by the whistlers, which then move to the right. The fluctuations move slowly compared to the thermal speed and most do not reach the cold reservoir at $x=L_{0}$ by the end of the simulation at $t=800 \: \Omega_{e0}^{-1}$. (b) Fourier spectra, summed over $k_{y}$ and plotted as a function of $k_{x}$ for the simulations with $L_{x}=L_{0}$ and $\beta_{e0h}=32$, $64$, and $128$.}
    \label{fig:5}
\end{figure}
In figure \ref{fig:5}a we show a spacetime diagram ($t$ versus $x$) of
the out-of-plane $B_{z}$ at a single value of $L_{y}/2$. After a
transient associated with the anisotropy-driven waves of the initial
distribution $f_{0}$ that was discussed earlier, the whistlers
propagate at a nearly uniform speed in the direction of $-T'$
($+\mathbf{\hat{x}}$). To confirm that the unstable modes have $k
\rho_{e} \sim 1$, we show the power spectrum $|B_{k_x}|^{2}$ for the
runs with $L_{x}=L_{0}$ at $\beta_{e0h}=32,64$ and $128$ in
fig. \ref{fig:5}b. The spectra are nearly isotropic in the 2D Fourier
space $k_x-k_y$ (not shown) so in the spectra shown the energy has
been summed over $k_{y}$. We find a spectral index of $-13/3$ for the
modes near $k_x\rho_{e0h}=1$ although we note that the more important
point is to establish that the spectrum peaks near $k
\rho_{e0h}=1$ even as $\beta_{e0h}$ varies. A more complete exploration
of the spectrum requires simulations with a third spatial
dimension. In addition we find that for each of the six simulations in
Table 1, $q_{ex,f} \simeq 3 \: n_0 v_{p} T_{eh}$, where $v_{p}$ was
measured in the middle of the simulation domain. These results
strongly support the scaling
\begin{equation} \label{eqn:3}
    q_{\parallel} = \alpha n_{0} \frac{\omega}{k} T_{eh} \sim n_{0} \frac{v_{Teh}^{3}}{\beta_{e0h}} = v_{Teh} \frac{B_{0}^2}{2},
\end{equation}
where $\alpha$ is a coefficient of order unity. Equation (\ref{eqn:3})
reveals the crucial role of the background magnetic field in
facilitating thermal transport since it controls the propagation of
whistlers. In the case of a very small magnetic field the whistlers
barely propagate and the thermal conduction is virtually shut
off. However, no whistler growth was found in a simulation with
$B_{0}=0$ (not shown), indicating that heat flux suppression by
whistlers requires a finite ambient magnetic field. Recent PIC
simulations with an imposed thermal gradient suggest that pressure
anisotropy driven modes are at play when there is no initial ambient
magnetic field \cite{Schoeffler2017}. Those results are consistent
with the transient growth of fluctuations seen in our simulations in
the case of $B_{0}=0$. These reach finite amplitude but then rapidly
decay on time scales short compared with the development of the
heat-flux instability.

\textit{Discussion}.  A caveat of our model is that the imposed
thermal gradient is much larger than that measured in environments
such as the ICM \cite{Levinson1992}. However, the present simulations
suggest that the transport is insensitive to the imposed temperature
gradient (although the sign of the parallel heat flux is determined by
the sign of $-\nabla T_{e}$ through the whistler phase speed). The
point is that heat flux instability is directly driven by the
collisionless heat flux, which depends only on the temperature
difference across a domain, rather than the ambient gradient. It seems
likely, therefore, that the current results apply to cases in which
the temperature gradient is far weaker. A full treatment of the ICM
also requires the inclusion of weak collisions not present in our
kinetic model.

A question is how the microphysics of whistler scattering will affect
heating and thermal conduction in the intracluster medium. The scaling
of heat flux in (\ref{eqn:3}) with $1/\beta_e$ implies a suppression
factor of roughly $100$ below the free-streaming thermal conduction.
The functional dependence $q_{\parallel} \propto T^{1/2}$ is a
noticeable departure from the Spitzer conductivity \cite{Spitzer1962}
proportional to $T^{7/2}$ often used in hydrodynamic or MHD models of
the ICM (e.g. \cite{Ruszkowski2010}, \cite{Yang2016b}). Our results
may therefore significantly alter the equilibria associated with
clusters of galaxies, which result from a balance between thermal
conduction and radiative cooling.

Our results show promising similarities with the observations of
thermal conduction in the solar wind by Bale et al. \cite{Bale2013} in
which the heat flux takes on a constant value, independent of
collisionality and the ambient temperature gradient, in the weak
collisionality regime where the collisional mean-free-path exceeds the
temperature scale length. However, much of their data is in a regime
of much lower $\beta$ than in the present simulations. The exploration
of the transition from high to low $\beta$ with analysis and
simulations is underway so that more detailed comparisons with solar
wind observations can be made.

\bibliography{library}
%\begin{thebibliography}
%\end{thebibliography}

\end{document}

%\preprint{APS/123-QED}

\title{Supplementary Material for: Suppression of electron thermal conduction by whistler turbulence in a sustained thermal gradient}

\author{G. T. Roberg-Clark}
\email{grc@umd.edu}
\affiliation{Department of Physics, University of Maryland College Park, College Park, MD 20740, USA}
\author{J. F. Drake}%
\email{drake@umd.edu}
\affiliation{Department of Physics, University of Maryland College Park, College Park, MD 20740, USA}
\affiliation{Institute for Physical Science and Technology, University of Maryland, College Park, MD 20742, USA}
\affiliation{Institute for Research in Electronics and Applied Physics, University of Maryland, College Park, MD 20742, USA}
\affiliation{Joint Space-Science Institute (JSI), College Park, MD 20742, USA}
\author{C. S. Reynolds}%
\email{chris@astro.umd.edu}
\affiliation{Department of Astronomy, University of Maryland College Park, College Park, MD 20740, USA}
\affiliation{Joint Space-Science Institute (JSI), College Park, MD 20742, USA}
\author{M. Swisdak}%
\email{swisdak@umd.edu}
\affiliation{Department of Physics, University of Maryland College Park, College Park, MD 20740, USA}
\affiliation{Institute for Research in Electronics and Applied Physics, University of Maryland, College Park, MD 20742, USA}
\affiliation{Joint Space-Science Institute (JSI), College Park, MD 20742, USA}

\date{\today}

\maketitle

\textit{Isotropization of the distribution function}. Here we demonstrate strong scattering of the electron distribution function for the large simulation quoted in the main paper with $L_{x}=4 L_{0}$, $\beta_{e0h}=64$. The distribution function is sampled from a thin band around the center of the domain at $L=L_{x}/2$, averaged over all $y$. In fig. \ref{fig:S1}a the initial (highly anisotropic) $f_{0}$ at $t=0$ is shown in $v_{x} - v_{y}$ space, with $v_{z}$ dependence integrated out. By the end of the simulation (fig. \ref{fig:S1}b) the electron distribution function is much more isotropic. This is accomplished via the resonant overlap mechanism mentioned in the main paper.

\renewcommand{\thefigure}{S\arabic{figure}}

\begin{figure}[b]
    \centering
    \includegraphics[scale=.55]{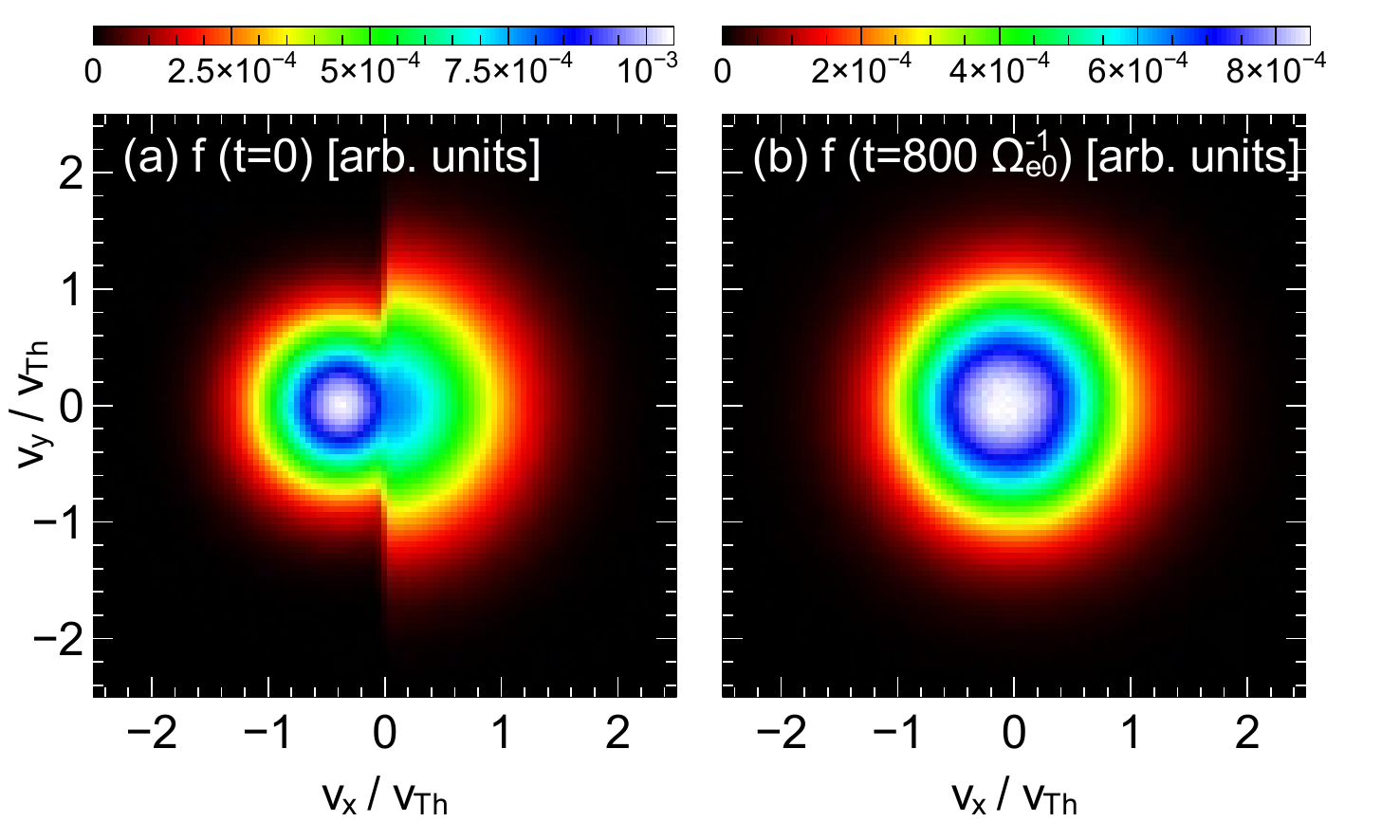}
    \caption{Evidence for isotropization of the distribution function by whistler scattering by late time in the center of the simulation domain. (a) $f(t=0)$ as a function of $v_{x}$ and $v_{y}$. (b) $f(t=800\Omega_{e0}^{-1})$ }
    \label{fig:S1}
\end{figure}